\documentclass[aps,twocolumn,nofootinbib,preprintnumbers]{revtex4-1}

\usepackage{amsmath,amssymb,amsfonts,amsthm,graphicx, epsf,dsfont,dcolumn,yfonts}
\usepackage{latexsym}
\usepackage[colorlinks=true, pdfstartview=FitV, linkcolor=black, citecolor=black, urlcolor=black]{hyperref}

\usepackage{color}
\usepackage{slashed} 

\newcommand{\be}{\begin{equation}}
\newcommand{\ee}{\end{equation}}
\newcommand{\bea}{\begin{eqnarray}}
\newcommand{\eea}{\end{eqnarray}}

\newcommand{\bwt}{\begin{widetext}}
\newcommand{\ewt}{\end{widetext}}

\def\G{\Gamma}

\def\ads{{\rm AdS}}

\begin{document}

\title{Dynamically Generated Gap from Holography: Mottness from a
  Black Hole}

\author{Mohammad Edalati, Robert G. Leigh and Philip W. Phillips}
\affiliation{ Department of Physics,
University of Illinois at Urbana-Champaign, Urbana IL 61801, USA}

\begin{abstract}
In the fermionic sector of  top-down approaches to holographic systems, one generically finds that the fermions are coupled to
gravity and gauge fields in a variety of ways, beyond minimal coupling. In this paper, we take one such interaction -- a Pauli, or dipole, interaction -- and study its effects on fermion correlators. We find that this interaction modifies the fermion spectral density in a remarkable way. As we change the strength of the interaction, we find that spectral weight is transferred between bands, and beyond a critical value, a gap emerges in the fermion density of states. A possible interpretation of this bulk interaction then is that it drives the dynamical formation of a (Mott) gap, in the absence of continuous symmetry breaking.
\end{abstract}

\maketitle


Holography provides a method to study the dynamics of certain strongly correlated systems. In the recent past, there has been much discussion about
the two-point correlation functions (and thus the spectral density) of probe fermionic operators coupled to these strongly correlated systems; see for example \cite{Lee:2008xf, Liu:2009dm, Cubrovic:2009ye, Faulkner:2009wj
} and their citations. By modifying the geometry and field content of the bulk geometry, one changes the properties of the strongly coupled system under study. The infra-red properties of the correlation functions of the probe operators are determined both by the properties of the dual geometry and how the probes are coupled to it. A variety of emergent infra-red properties have been identified.

In most cases, we do not possess a microscopic understanding of the field theory dynamics under study, beyond simple properties such as temperature, charge density, etc. In the so-called ``bottom-up" approaches, one modifies these simple properties by changing features of the bulk theory. In some situations, where we have explicit `ultra-violet completions' in terms of string or M-theory constructions, one may know more of how to interpret the dual field theory. Starting in ten or eleven dimensions, one can obtain (super)gravitational descriptions of the bulk physics, and in some cases corresponding knowledge of the dual field theory. Recently, it was shown \cite{Gauntlett:2009zw,Cassani:2010uw,Gauntlett:2010vu,Liu:2010sa} that there are consistent truncations of ten and eleven dimensional supergravities to five and four dimensional bulk theories that possess an interesting class of gauge interactions and charged matter. 
The gauge and matter content allows for interesting condensed matter physics phenomena in either two or three (spatial) dimensions, such as superconductivity, to be explored in a fully consistent `top-down' approach \cite{Denef:2009tp, Gubser:2009qm, Gauntlett:2009dn}. In addition to this construction, there are of course many other brane constructions that have and can be used. 

Recently, the fermionic content of these consistent truncations has been worked out in detail \cite{Bah:2010yt,Bah:2010cu}. We emphasize here that the fermionic sectors possess a number of generic features, many of which have not been explored in the holographic literature to date, and thus there is a good chance that these features might lead to interesting phenomena related to condensed matter physics. In this paper, we study a simple example in which a fermion is coupled to a gauge field through a dipole interaction in the bulk. Remarkably, we find that as the strength of this interaction is varied, a new band in the density of states emerges, spectral density is transferred between bands, and, beyond a critical interaction strength, a gap opens up.\footnote{Our analysis here is in contrast to the one carried out on a holographic superconducting background \cite{Faulkner:2009am} where by coupling a probe fermion to the charged scalar $\phi$ of the bulk through the so-called Majorana coupling $\eta_5^*\phi^* \psi^T C \,\Gamma^5 \psi+h.c.$ (with $\eta_5$ and $C$ being the strength of the Majorana coupling and  the charge conjugation matrix, respectively), a gap to the creation of quasiparticle-hole pairs was observed in the spectral density of the dual fermionic operator near the Fermi surface. Such a gap is a result of coupling to a symmetry breaking condensate (where the probe fermion is coupled to the vev of the scalar through the aforementioned Majorana coupling) and is distinct from that studied here.}

Such phenomena occur in several important condensed matter systems. The typical systems, Mott insulators, where a gap is dynamically generated, possess a half-filled band \cite{mott, phillips_rmp}.
A key consequence
of the dynamical generation of a charge gap is that of {\it spectral weight
transfer} \cite{sawatzky, chen} --- the removal of a single fermionic charge carrier (attained by doping)
results in rearrangements of the spectrum on all energy scales.  Since
the formation of bound states \cite{mott,mott_gap,phillips_rmp} in the Mott problem represents an
example of strong-coupling physics it should in principle be
describable by gauge-gravity duality\footnote{The idea of modeling Mott insulators using holography has been previously mentioned in the literature \cite{Hartnoll:2007ih,Balasubramanian:2010uw}, although in a schematic way and not in regard to fermion correlators.}.  The holographic system that we study in this paper describes both the dynamical appearance of a gap to low energy fermion transport and spectral weight transfer, the two key features of Mott physics. 


The consistent truncation referred to above gives rise to a number of generic couplings of fermions. One feature which we believe to be important, but will not consider here, is the fact that quite generically one finds couplings between spin-$1/2$ and spin-$3/2$ fermions.  Since string or M-theory is at its core a gravitational theory with supersymmetry, spin-$3/2$ gravitinos (massive or massless) are generically present and coupled to other fermions. 
In addition, there are gauge fields and both charged and uncharged matter fields, such as scalars, that are coupled in specific ways, and, depending on the range of parameters, could cause instabilities. For applications to holographic superconductors, one finds suitable Majorana charge couplings for example.

 In this paper, we consider just one form of non-minimal coupling, in which the gauge field couples to spin-$1/2$ fermions through a dipole interaction of the form $F_{ab}\bar\psi \Gamma^{ab}\psi$. 
In fact, we will only consider here the simplest possible setup, in which the fermion propagates in the background of a Reissner-Nordstr\"om (RN) AdS$_{d+1}$  black-hole. In this sense, we are employing a bottom-up construction, inspired by generic top-down models. It would be interesting to extend our results to other bulk systems, for example those in which there are also charged scalars which can, in principle, condense and give rise to superconductivity.


Thus, we consider here the Lagrangian (in $d+1\geq4$ dimensions)
\begin{align}
\sqrt{-g}\,i\bar\psi (\slashed{D}-m- ip \slashed{F})\psi,
\end{align}
where $\slashed{D}=e_c^M\G^c\left(\partial_M+\frac{1}{4}\omega_M^{~ab}\G_{ab}-iqA_M\right)$ 
and  $\slashed{F}=\frac{1}{2}\G^{ab} e_a^{M}e_b^{N}F_{MN}$, with $e^M_a$ and $\omega_M^{~ab}$ being the (inverse) vielbein and the spin connection, respectively.

For notational simplicity, we will rescale $p$, namely $p\to p L/(d-2)$, in what follows. 
The significance of the parameters $m$ and $q$ are well-known: they correspond to the scaling dimension and global $U(1)$ charge of dual fermionic operators. The significance of $p$ in the dual field theory is not apparent however. In the holographic system, since $p$ is a coupling of the probe fermion to the bulk geometry, we expect that it may modify the correlation functions of the dual fermionic operator.

The Reissner-Nordstr\"om AdS$_{d+1}$ (RN-AdS) geometry has a metric and a gauge connection which can be written
\bea
ds^2&=&\frac{r^2}{L^2}\left[- f(r) dt^2 + d\vec x^2 \right]+\frac{L^2}{r^2} \frac{dr^2}{f(r)},\\
A&=&\mu[1-(r_0/r)^{d-2}]dt,
\eea
where $f(r)=1-M\left(r_0/r\right)^d+Q^2\left(r_0/r\right)^{2(d-1)}$, $M=1+Q^2$,  
$\mu=\sqrt{(d-1)/(2d-4)}\,Qr_0/L^2$, with $r_0$ being the horizon radius.
To decouple the equation of motion, we introduce projectors $\Gamma_\pm=\frac{1}{2}(1\pm \Gamma^{\underline r}\Gamma^{\underline t}\hat k\cdot\vec\Gamma)$ and write $\psi_\pm(r)=r^{d/2} f(r)^{1/4}\Gamma_\pm\psi(r)$. Without loss of generality, we take $k_1=k$ and $k_{i\neq 1}=0$, and take the basis 
\bea\label{GammaBasis}
 \Gamma^{\underline r} &=& \begin{pmatrix}
   -\sigma_3 \otimes\mathds{1} & 0 \\
    0&   -\sigma_3 \otimes\mathds{1}
   \end{pmatrix}, \hskip 0.03in\Gamma^{\underline t} = \begin{pmatrix}
   i\sigma_1 \otimes\mathds{1} & 0 \\
    0&   i\sigma_1 \otimes\mathds{1}
   \end{pmatrix},\nonumber\\ \Gamma^{\underline 1} &=& \begin{pmatrix}
   -\sigma_2 \otimes\mathds{1} & 0 \\
    0&   \sigma_2 \otimes\mathds{1}
   \end{pmatrix}.
\eea
One then finds
\begin{widetext}\bea\label{DiracEqInRN}
\left(\frac{r^2}{L^2}\sqrt{f(r)}\,\partial_r+\frac{r}{L}m\,\sigma_3\right)\psi_{\pm}=
\left[\frac{i\sigma_2}{\sqrt{f(r)}}\left[\omega
  +\mu q
  \left(1-\frac{r_0^{d-2}}{r^{d-2}}\right)\right]-\left(\mu p\frac{r_0^{d-2}}{r^{d-2}}\pm k\right) \sigma_1\right]\psi_{\pm}.
\eea
\end{widetext}
We see that the principle effect of the Pauli coupling is to modify the appearance of $k$ in the equation of motion. As we will explain below, this feature can lead to a gap in the spectral density.

To see the effects more clearly, consider the solutions of the Dirac equations \eqref{DiracEqInRN} in the two (asymptotic and near horizon) regimes. Asymptotically, the solutions behave as 
\bea\label{psipmAsymptotics}
\hskip -0.02in\psi_{\pm}(r, \omega, k)= a_\pm (\omega,k)\,r^{mL} \begin{pmatrix}
0 \\ 1 \end{pmatrix}+b_\pm  (\omega,k)\,r^{-mL} \begin{pmatrix}
1 \\ 0 \end{pmatrix}\hskip -0.03in.\,\,
\eea
The effect of $p$ asymptotically is to modify the subleading terms. In this paper, we take $m\in [0,\frac{1}{2})$ and consider $a_{\pm}$ to be the sources (conventional quantization). Having chosen in-falling boundary condition near the horizon, the retarded correlator is then of the form 
\bea\label{ConventionalGreenFunctions}
G_R(\omega, k) = \begin{pmatrix}
   G_{+}(\omega, k)\, \mathds{1} & 0 \\
    0&      G_{-}(\omega, k)\, \mathds{1}
   \end{pmatrix},
\eea
with $G_{\pm}(\omega, k) = b_{\pm}(\omega, k)/a_{\pm}(\omega, k)$.
Note that the Dirac equations \eqref{DiracEqInRN} imply $G_+(\omega,k)=G_-(\omega,-k)$.
\begin{figure*}
 \begin{center}
\includegraphics[width=68mm]{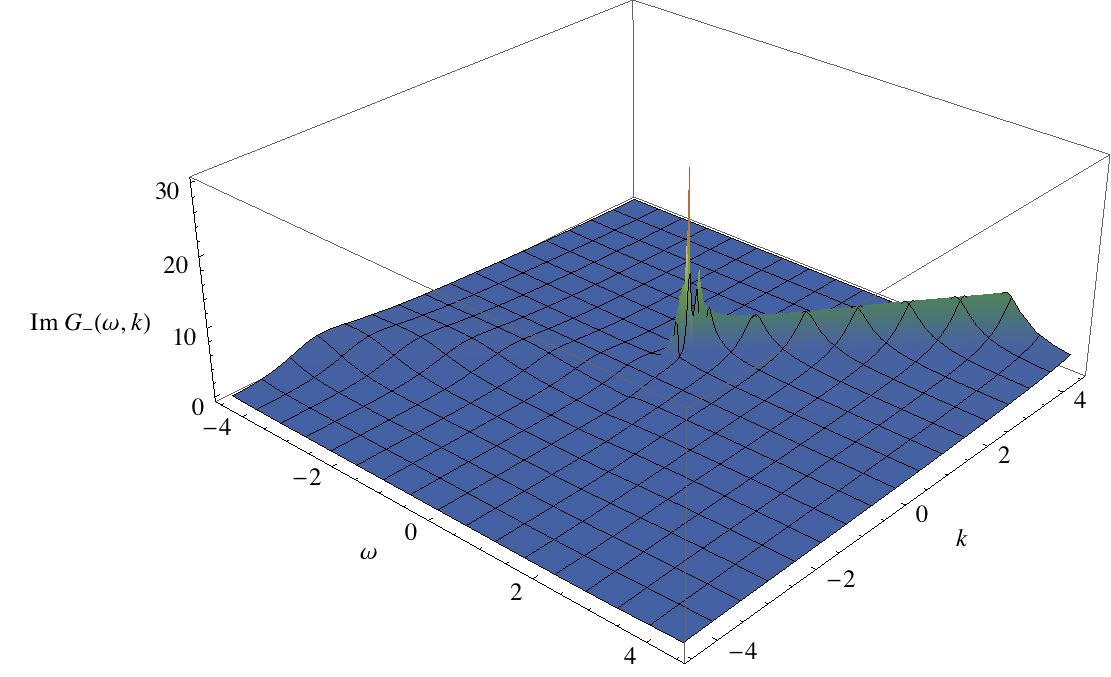}
\hskip0.2in
\includegraphics[width=67mm]{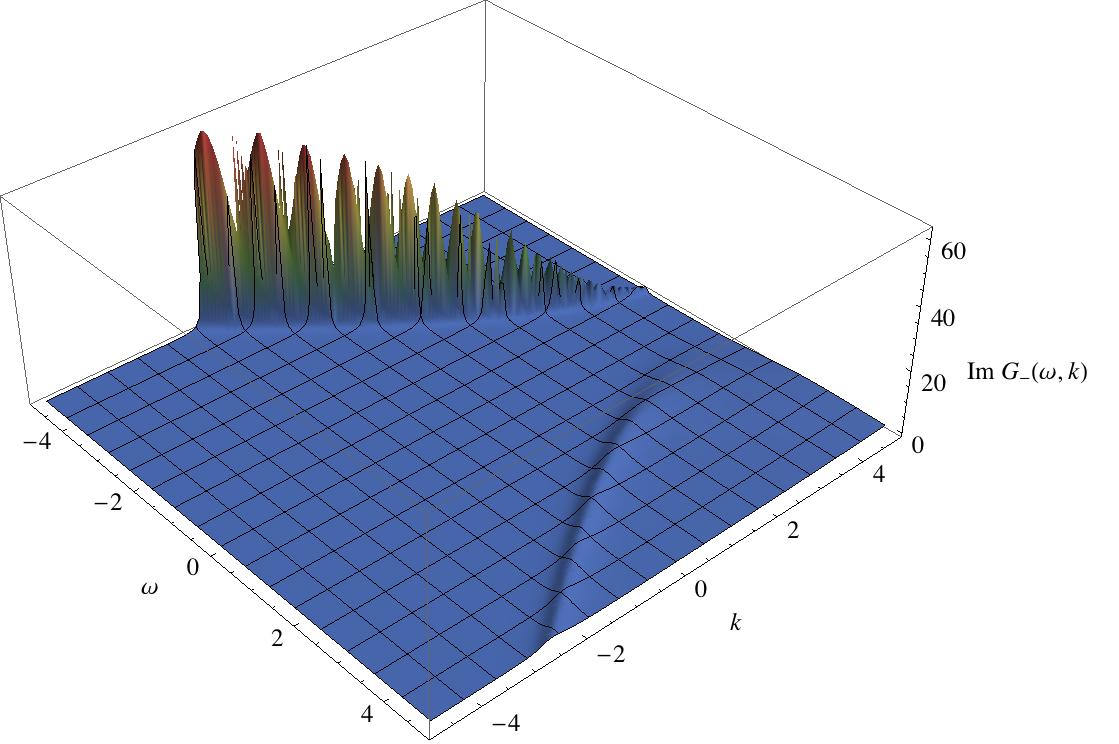}
\vskip-0.1in\caption{\footnotesize{${\rm Im}\,G_{-}(\omega,k)$ for $p=0$ and  $p=4.5$. A gap is clearly visible around $\omega=0$ in the second plot. We show just $G_-(\omega,k)$; $G_+(\omega,k)$ can be recovered using the relation $G_+(\omega,k)=G_-(\omega,-k)$ implied by the Dirac equation. We have set $L=r_0=1$.} }
\end{center}
\end{figure*}
In this note, we consider the extremal case (zero temperature). When the background is extremal, $f(r)$ has a double zero at the horizon, and this fact renders the limit $ \omega \to 0$ of the Dirac equations \eqref{DiracEqInRN} near the horizon subtle. 
To take care of the subtlety, one realizes \cite{Faulkner:2009wj,Edalati:2009bi} that near the horizon (in which the geometry approaches AdS$_2\times \mathds{R}^2$ for $T=0$) the equations for $\psi_{\pm}$ in \eqref{DiracEqInRN} organize themselves as functions of $\zeta = \omega L_2^2/(r-r_0)$ with $L_2= L/\sqrt{d(d-1)}$ being the radius of AdS$_2$. 
The coordinate $\zeta$ is the suitable radial coordinate for the $\ads_2$ part of the near horizon region, and in this region, we can write $\psi_{\pm}$ in terms of $\zeta$ and expand in powers of  $\omega$ as follows
\begin{align}\label{innerPhi}
\psi_{I\pm}(\zeta)&=\psi^{(0)}_{I\pm}(\zeta)+\omega\,\psi^{(1)}_{I\pm}(\zeta)+\omega^2 \psi^{(2)}_{I\pm}(\zeta)+\cdots.
\end{align}
Now, substituting \eqref{innerPhi} into \eqref{DiracEqInRN}, we find that to leading order
\bea\label{leadinglinner}
\psi^{(0)\prime\prime}_{I\pm}(\zeta)&=&\frac{L_2}{\zeta}\left[m\sigma_3+\left(c_d\frac{p}{L} \pm\frac{kL}{r_0}\right)\sigma_1\right]\psi^{(0)}_{I\pm}(\zeta)\nonumber\\
&&-\,i\sigma_2\left(1+\frac{q e_d}{\zeta}\right)\psi^{(0)}_{I\pm}(\zeta),
\eea
where $e_d=1/\sqrt{2d(d-1)}$ and $c_d=1/\left[(2d-4)e_d\right]$.  Equations \eqref{leadinglinner} are identical to the equations of motion for massive spinor fields \cite{Faulkner:2009wj} with masses $(m, \tilde m_+)$ and $(m,\tilde m_-)$ in AdS$_2$, where $\tilde m_\pm$ are time-reversal violating mass terms, with the identification
\begin{align}
\tilde m_\pm =c_d\frac{p}{L} \pm\frac{kL}{r_0}. 
\end{align}
Thus,  $\psi^{(0)}_{I\pm}(\zeta)$ are dual to spinor operators ${\cal O}_{\pm}$ in the IR CFT with conformal dimensions $\delta_{\pm}=\nu_{\pm}+\frac{1}{2}$ where 
\bea\label{nupm}
\hskip -0.04in\nu_{\pm}=\sqrt{m_{k\pm}^2L_2^2-q^2e_d^2-i\epsilon}, \,m_{k\pm}^2\hskip -0.05in=m^2 \hskip-0.05in+\hskip-0.03in\left(c_d\frac{p}{L}\pm\frac{kL}{r_0}\right)^2\hskip -0.1in . \,\,\,\,
\eea
We see that turning on $p$ modifies the scaling in the infrared in an important way -- effectively, the momentum is pushed up or down by $p$.

In what follows, we will numerically solve the Dirac equations \eqref{DiracEqInRN} for generic $\omega$ and $k$. It is convenient to write $\psi_{\pm}^T=(\beta_{\pm}, \alpha_{\pm})$ and define $ \xi_{\pm}=\beta_{\pm}/\alpha_{\pm}$, in terms of which the Dirac equations \eqref{DiracEqInRN} then reduce to flow equations
\bea\label{FlowEquations}
\hskip-0.3in\frac{r^2}{L^2}\sqrt{f(r)}\partial_r\xi_{\pm}&=&-2m\frac{r}{L}\xi_{\pm}\nonumber\\
&&+\left[v_{-}(r)\mp k\right]+\left[v_{+}(r)\pm k\right]\xi^2_{\pm}, \hskip 0.06in
\eea
where
\begin{align}
v_{\pm}(r)=\frac{1}{\sqrt{f(r)}}\left[\omega+\mu q \left(1-\frac{r_0^{d-2}}{r^{d-2}}\right)\right]\pm \mu p \left(\frac{r_0}{r}\right)^{d-2}.\nonumber
\end{align}
Expressed in terms of $\xi_{\pm}$, the matrix of Green functions \eqref{ConventionalGreenFunctions} becomes
\begin{align}\label{XiPmGreenFunctions}
G_R(\omega, k) = \lim_{\epsilon\to0}\,\epsilon^{-2mL}\begin{pmatrix}
   \xi_{+}\, \mathds{1} & 0 \\
    0&       \xi_{-}\, \mathds{1}
   \end{pmatrix}\Big|_{r=\frac{1}{\epsilon}},
\end{align}
where one is instructed to pick the finite terms as $\epsilon\to 0$.  


We focus on the $d=3$ case and consider the extremal RN-$\ads_{4}$ background (for which $Q^2=3$). Thus, we have a (2+1)-dimensional  boundary theory at zero temperature and finite charge density. Our results below are easily carried over to higher dimensions. We take $m=0$, so that the scaling dimension of the boundary theory fermion operator is $\Delta_\psi=3/2$. Also, in what follows, we set $q=1$ as  we vary $p$. The effect of non-zero $p$ for other values of the parameters $m$ and $q$ will be discussed elsewhere.
\begin{figure}
\centering
\begin{tabular}{ccc}
\hskip -0.15in\includegraphics[width=44mm]{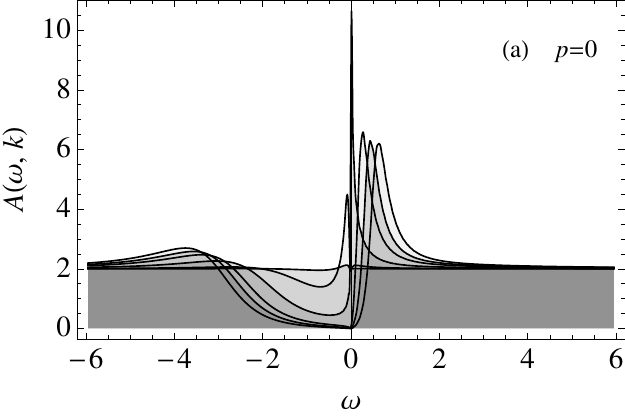} &
\includegraphics[width=44mm]{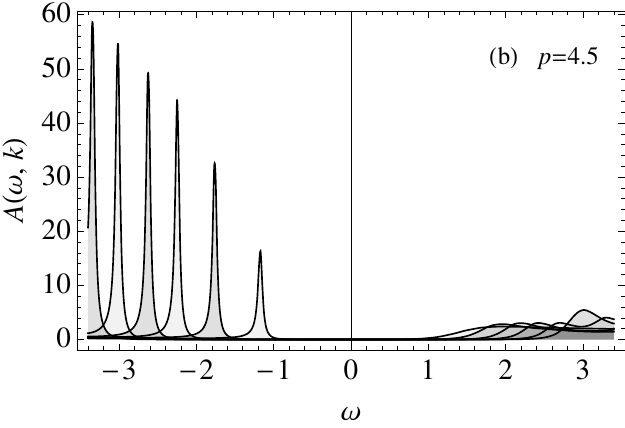} 
\end{tabular}\vskip-0.1in
\caption{\footnotesize{Figure 2(a) shows $A(\omega, k)$ as a function of $\omega$ for sample values of $k\in [0.1,3.2]$ for $p=0$. $k=k_F\simeq 0.92$ for the peak at $\omega=0$. Figure 2(b) shows the same plot but for $p=4.5$. The gap around $\omega=0$ persists for all values of $k$. } }\vskip-0.1in
\end{figure}
To compute the retarded Green function (and hence the spectral density) of the dual fermionic operator one imposes an in-falling boundary condition near the horizon for $\psi_{\pm}$ \cite{Iqbal:2009fd}. In terms of $\xi_\pm$, the in-falling boundary condition reads (for $\omega\neq0$) 
$\xi_{\pm}(r_0,\omega,k)=i.
$ We can now numerically integrate the flow equations \eqref{FlowEquations} and read off the asymptotic values of  $\psi_{\pm}$ from which the matrix of retarded Green functions is easily computed using \eqref{XiPmGreenFunctions}. The quantity of interest for us is the fermion spectral function which, up to normalization, is given by
$A(\omega,k) = {\rm Im }\left[{\rm Tr}\,G_R(\omega,k)\right]
$.  We will also be interested in the density of states $A(\omega)$, which is given by the integral of $A(\omega, k)$ over $k$. 

The plots in Figure 1 show  ${\rm Im}\,G_{-}(\omega,k)$ for $p=0$ and  $p=4.5$. The left plot has $p=0$,  analyzed previously in \cite{Liu:2009dm}: the quasi-particle-like peak represents a Fermi surface ($k_F\simeq\pm0.92$) where the excitations near the surface are of non-Fermi liquid type. The right plot, for which $p=4.5$,  clearly shows a gap around $\omega=0$. In Figure 2, we plot $A(\omega,k)$ for $p=0$ and $p=4.5$ for sample values of $k$. Figure 2(b) emphasizes that the gap in the spectral density exists for all $k$. 
In Figure 3, we show the density of states near the chemical potential
for various values of $p$. Our numerical computations indicate that
the onset of the gap is  near $p=4$. Finally, in Figure 4, we plot the width of the gap $\Delta $ versus $p$.
\begin{figure}
\includegraphics[width=47mm]{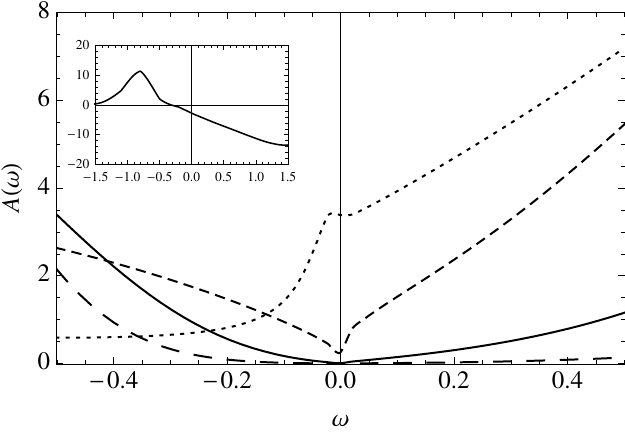} 
\vskip -0.17in\caption{\footnotesize{$A(\omega)$ for  $p=0$ (dotted), 2 (dashed),  4 (solid) and 6 (large dashed). The onset of the gap is at $p\simeq4$. The inset shows $A(\omega, p=6)-A(\omega,p=0)$ as a function of $\omega$.} }
\end{figure}
\begin{figure}
\hskip-0.1in\includegraphics[width=47mm]{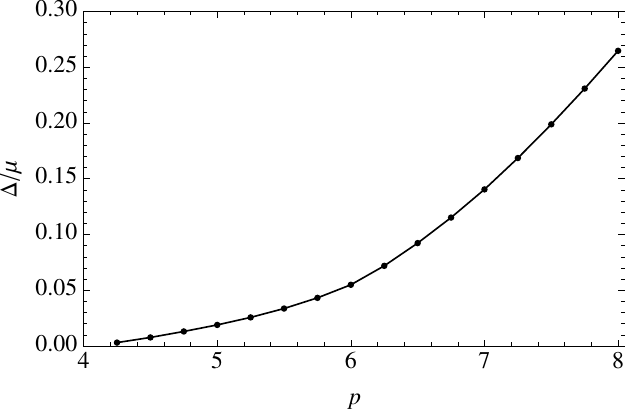} 
\vskip -0.17in\caption{\footnotesize{The width $\Delta$ (in units of $\mu$) of the dynamically generated gap as a function of $p$. } } \vskip-0.1in
\end{figure}
What we see from the numerical results is the following. For small
$p$, the dominant feature of the spectral density is the well-known
Fermi peak at $k=k_F$. As $p$ increases, the intensity of this peak
degrades, and spectral density begins to appear at negative
$\omega$. We will refer to this as the lower band. At a critical value
$p_{\rm crit.}$, a gap separating the lower band from the original
(upper) band emerges, {\it for all $k$}. As $p$ increases further,
this gap widens, and peaks begin to appear in the lower band. The
difference between $A(\omega,p=0$ and $A(\omega,p=6)$ is shown in the
inset of Fig. 3.  This illustrates clearly that the gap is accompanied
by transfer of spectral weight over all energy scales, not just on the
order of the gap as typified by systems in which a continuous symmetry
is broken for example BCS superconductors. 

The behavior for $p>p_{\rm crit.}$ strongly reminds us of the Mott gap
in a half-filled band in which no continuous symmetry is broken in the
formation of the gap.  This would naively suggest that $p$ plays the
role of the dimensionless interaction strength, $U/t$, in the Hubbard
model. However, in the present work lowering $p$ not only closes the
gap but also shifts spectral weight from the upper to the lower band
(Fig. 3 inset).  In the Hubbard model, it is the doping that leads to spectral weight transfer.  Hence, the parameter $p$ seems to, in terms of the Hubbard model, mimic the combined effects of doping and interaction strength.
It would be interesting to determine (but is beyond the scope of this discussion) if a condensate violating some discrete symmetry is present 
when $\Delta\neq0$. Note however, 
that such a transition is expected to  only involve energy scales 
$O(\Delta)$, in contrast to the re-distributions of the spectral weight on all energy  scales seen in Figure 3.

Finally, it is important to appreciate the vagaries of holographic studies such as this one. We are not claiming that turning on the Pauli coupling to a probe fermion opens up a gap in charge transport. Indeed, because we are not considering back reaction in any sense, one expects that charge current correlators are unmodified. The probe fermion makes a negligible contribution to charge transport in the dual field theory. However, transport of the dual fermion operator is gapped.

\section*{Acknowledgments}
We would like to thank E. Fradkin, T. Hughes and J. Jottar for discussions.  E. M. and P. P.  acknowledge financial support from the NSF DMR-0940992 and the Center for Emergent Superconductivity, a DOE Energy Frontier Research Center, Award Number DE-AC0298CH1088. R.G.L.\ is  supported by DOE grant FG02-91-ER40709.


\end{document}